\pdfoutput=1
\documentclass[11pt,twoside,a4paper,cmspaper,final,collab]{cms-tdr}
\def\svnVersion{8809407}\def\svnDate{2024/12/20}\def\cmsCernNoTag{CERN-EP-2024-123}\def\cmsCernDate{\today}\def\cmsMessage{Published in Physics Letters B as \href{http://dx.doi.org/10.1016/j.physletb.2024.139202}{\doi{10.1016/j.physletb.2024.139202}.}}
\begin{document}\cmsNoteHeader{HIG-23-007}

\newcommand{\Pj}{\ensuremath{\mathrm{j}}\xspace}

\newcommand{\kW}{\ensuremath{\kappa_{\PW}}\xspace}
\newcommand{\kZ}{\ensuremath{\kappa_{\PZ}}\xspace}
\newcommand{\lambdaWZ}{\ensuremath{\lambda_{\PW\PZ}}\xspace}

\newcommand{\HWW}{\ensuremath{\PH{}\PW{}\PW}\xspace}
\newcommand{\HZZ}{\ensuremath{\PH{}\PZ{}\PZ}\xspace}
\newcommand{\WH}{\ensuremath{\PW{}\PH}\xspace}
\newcommand{\Htobb}{\ensuremath{\PH\to\bbbar}\xspace}
\newcommand{\Xtobb}{\ensuremath{X\to\bbbar}\xspace}
\newcommand{\Wjets}{\ensuremath{\PW+\text{jets}}\xspace}
\newcommand{\Zjets}{\ensuremath{\PZ+\text{jets}}\xspace}

\newcommand{\MSD}{\ensuremath{m_{\mathrm{SD}}}\xspace}
\newcommand{\Mjj}{\ensuremath{m_{\Pj{}\Pj}}\xspace}
\newcommand{\Mbb}{\ensuremath{m_{\PQb{}\PQb}}\xspace}
\newcommand{\detajj}{\ensuremath{\Delta\eta_{\Pj\Pj}}\xspace}
\newcommand{\absdetajj}{\ensuremath{\abs{\detajj}}\xspace}
\newcommand{\std}{\ensuremath{\,\sigma}\xspace}
\newcommand{\DeepJet}{\textsc{DeepJet}\xspace}
\newcommand{\ParticleNet}{\textsc{ParticleNet}\xspace}
\newcommand{\muF}{\ensuremath{\mu_{\mathrm{F}}}\xspace}
\newcommand{\muR}{\ensuremath{\mu_{\mathrm{R}}}\xspace}

\cmsNoteHeader{HIG-23-007}

\title{Study of \texorpdfstring{$\PW\PH$}{WH} production through vector boson scattering and extraction of the relative sign of the \PW and \PZ couplings to the Higgs boson in proton-proton collisions at \texorpdfstring{$\sqrt{s} = 13\TeV$}{sqrt(s) = 13 TeV}}

\date{\today}

\abstract{
A search for the production of a \PW boson and a Higgs boson through vector boson scattering (VBS) is presented, using CMS data from proton-proton collisions at $\sqrt{s} = 13\TeV$ collected from 2016 to 2018.  The integrated luminosity of the data sample is 138\fbinv. Selected events must be consistent with the presence of two jets originating from VBS, the leptonic decay of the \PW boson to an electron or muon, possibly also through an intermediate \PGt lepton, and a Higgs boson decaying into a pair of \PQb quarks, reconstructed as either a single merged jet or two resolved jets. A measurement of the process as predicted by the standard model (SM) is performed alongside a study of beyond-the-SM (BSM) scenarios. The SM analysis sets an observed (expected) 95\% confidence level upper limit of 14.3 (9.9) on the ratio of the measured VBS WH cross section to that expected by the SM. The BSM analysis, conducted within the so-called $\kappa$ framework, excludes all scenarios with $\lambdaWZ < 0$ that are consistent with current measurements, where $\lambdaWZ = \kW/\kZ$ and \kW and \kZ are the \HWW and \HZZ coupling modifiers, respectively. The significance of the exclusion is beyond 5 standard deviations, and it is consistent with the SM expectation of $\lambdaWZ = 1$. 
}

\hypersetup{
pdfauthor={CMS Collaboration},
pdftitle={Study of WH production through vector boson scattering and extraction of the relative sign of the W and Z couplings to the Higgs boson in proton-proton collisions at 13 TeV},
pdfsubject={CMS},
pdfkeywords={CMS, Higgs, vector boson scattering}}

\maketitle 

\section{Introduction}
Precise measurements of the Higgs boson couplings to the \PW and \PZ bosons serve as critical tests of electroweak symmetry breaking in the standard model (SM)~\cite{Englert:1964et,Higgs:1964ia,Higgs:1964pj,Guralnik:1964eu,Higgs:1966ev,Kibble:1967sv}. 
Within the $\kappa$ framework~\cite{kframe}, where the \HWW and \HZZ coupling modifiers are denoted as \kW and \kZ, respectively, direct measurements of Higgs boson production and decay have tightly constrained the absolute values of \kW and \kZ~\cite{NatureHiggsCMS2022, NatureHiggsATLAS2022}. 
However, the sign of $\lambdaWZ = \kW/\kZ$ has not been probed directly.

\begin{figure*}[htb!]
    \centering
    \includegraphics[width=0.31\textwidth]{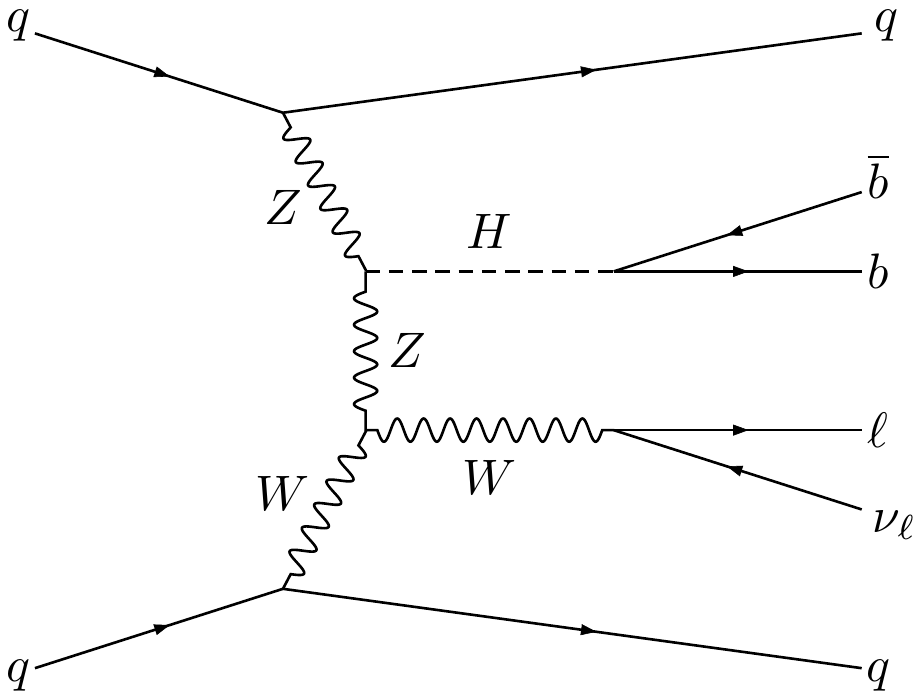}
    \includegraphics[width=0.31\textwidth]{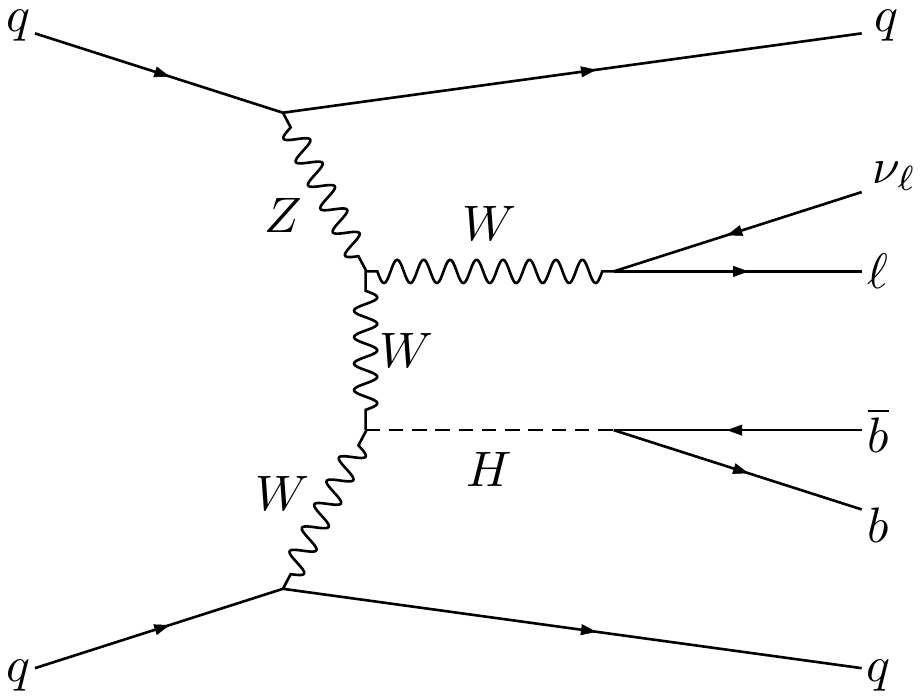} 
    \includegraphics[width=0.31\textwidth]{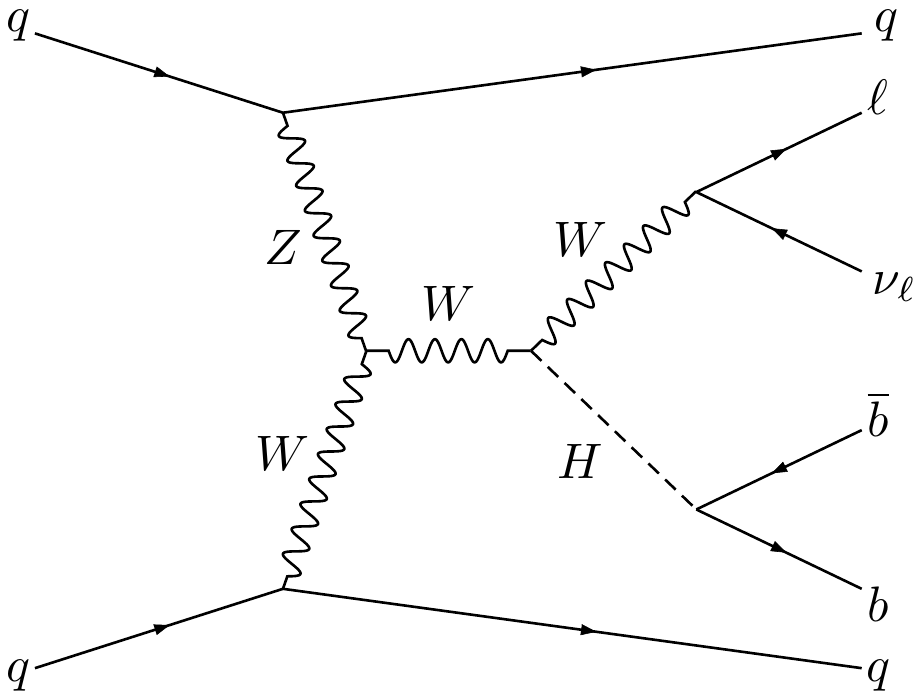} 
    \caption{
        Tree-level Feynman diagrams for the production of $\PW\PH\to\ell\PGn\PQb\PAQb$ via VBS, where $\ell$ is an electron or muon.
    }
    \label{fig:feynman_diagrams}
\end{figure*}

\begin{figure}[htb]
    \centering
    \includegraphics[width=0.49\textwidth]{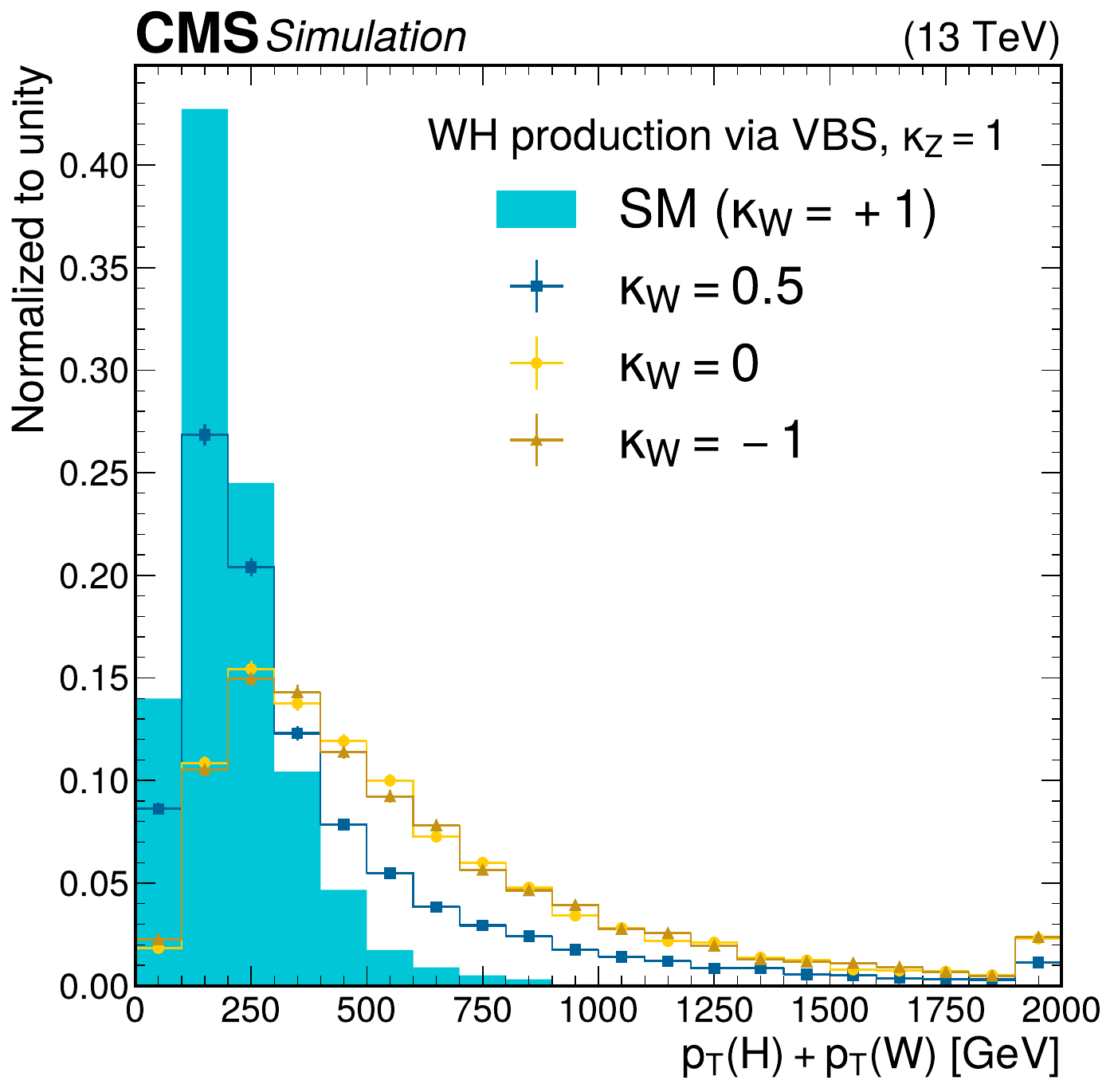}
    \caption{
        Distributions of the scalar sum of the Higgs boson and \PW boson \pt, as generated by \MADGRAPH, for simulated signal events with $\kZ=1$ and the following \kW values: $\kW=1$ (SM) in light blue (filled), $\kW=0.5$ in dark blue, $\kW=0$ in yellow, and $\kW=-1$ in gold. The highest bin also contains the overflows.
    }
    \label{fig:lhe_st}
\end{figure}

A process sensitive to the interference between the \HWW and \HZZ vertices is needed to extract the sign of \lambdaWZ.
A natural choice is the production of a W boson and a Higgs boson (\WH) via vector boson scattering (VBS)~\cite{Theory2LambdaWZ}, which proceeds primarily through the leading-order (LO) Feynman diagrams shown in Fig.~\ref{fig:feynman_diagrams}. 
This is a rare SM process, but the cross section grows quadratically with \kW or \kZ and the kinematic distributions are different for the SM ($\lambdaWZ = 1$), and $\lambdaWZ < 0$, where the \PW and Higgs bosons both receive a significant Lorentz boost. This is shown in Fig.~\ref{fig:lhe_st}, where the scalar sum of the transverse momentum (\pt) of the Higgs and \PW bosons is displayed for \kZ = 1 and various values of \kW.

In this Letter, we present a study of the production of \WH through VBS using the leptonic decay of the \PW boson to an electron or a muon, possibly also through an intermediate \PGt lepton, and the decay of the Higgs boson to \bbbar. 
The study focuses both on the SM scenario, where $\kW = \kZ = 1$, and on beyond-the-SM (BSM) scenarios, where $\lambdaWZ < 0$. 
The BSM analysis targets events where the Higgs boson decay to \PQb quarks is reconstructed as a single merged jet because of the Lorentz boost it receives, whereas the SM analysis targets events where the \PQb quarks from the Higgs boson are reconstructed as two resolved jets. 
The two scenarios are analyzed separately and use different methods. 
The study is based on data from proton-proton collisions at a center-of-mass energy of 13\TeV. 
The data were collected with the CMS detector~\cite{CMS:2008xjf} at the CERN LHC from 2016 to 2018, corresponding to an integrated luminosity of 138\fbinv. 
Tabulated results are provided in the HEPData record for this work~\cite{hepdata}.

\section{The CMS detector and object reconstruction}
The CMS apparatus~\cite{CMS:2008xjf,CMS:2023gfb} is a multipurpose, nearly hermetic detector, designed to trigger on~\cite{CMS:2020cmk,CMS:2016ngn} and identify electrons, muons, photons, and (charged and neutral) hadrons~\cite{CMS:2020uim,CMS:2018rym,CMS:2014pgm}. 
A global ``particle-flow'' (PF) algorithm~\cite{CMS:2017yfk} aims to reconstruct all individual particles in an event (PF candidates), combining information provided by the all-silicon inner tracker and by the crystal electromagnetic and brass-scintillator hadron calorimeters (ECAL and HCAL, respectively), operating inside a 3.8\unit{T} superconducting solenoid, with data from the gas-ionization muon detectors embedded in the flux-return yoke outside the solenoid. The PF candidates are used to build \PGt leptons, jets, and missing transverse momentum~\cite{CMS:2018jrd,CMS:2016lmd,CMS:2019ctu}. Forward calorimeters, made of steel and quartz-fibers, extend the pseudorapidity coverage provided by the barrel and endcap detectors. More detailed descriptions of the CMS detector, together with a definition of the coordinate system used and the relevant kinematic variables, can be found in Refs.~\cite{CMS:2008xjf,CMS:2023gfb}. 

Events of interest are collected using single-lepton triggers, where the charged lepton is required to be isolated from other significant detector activity and have \pt exceeding a threshold that depends on the data-taking period (year).
The single-muon trigger requires $\pt> 24\,(27)\GeV$ in 2016 and 2018 (2017), whereas the single-electron trigger requires $\pt > 27\,(32)\GeV$ in 2016 (2017--18). 

Muon candidates are reconstructed within the pseudorapidity range $\abs{\eta} < 2.4$ by combining the information from the silicon tracker and the muon chambers~\cite{CMS:2018rym}. The muon candidates are also required to satisfy a set of quality criteria based on the number of hits measured in the silicon tracker and the muon system, the properties of the muon track, and the impact parameters of the track with respect to the primary vertex (PV) of the event, which is taken to be the vertex corresponding to the hardest scattering in the event, evaluated using tracking information alone, as described in Section 9.4.1 of Ref.~\cite{CMS-TDR-15-02}. 

Electron candidates within $\abs{\eta} < 2.5$ are reconstructed by associating fitted tracks in the silicon tracker with electromagnetic energy clusters in the ECAL~\cite{CMS:2015xaf}. 
Electron candidates are required to satisfy identification criteria based on the shower shape of the energy deposit, the matching of the electron track to the ECAL energy cluster, the relative amount of energy deposited in the HCAL detector, and the consistency of the electron track with the PV. 
Electron candidates in the transition region between the ECAL barrel and endcaps, $1.44<\abs{\eta}<1.57$, are discarded because of suboptimal detector performance. 
Electron candidates identified as originating from photon conversions in the detector are also rejected.

Muons and electrons are required to be isolated from hadronic activity in the event. 
The isolation value is defined by summing the \pt of all PF candidates originating from the PV in a cone of radius $\Delta R = \sqrt{\smash[b]{(\Delta\eta)^2+(\Delta\phi)^2}} = 0.4\,(0.3)$ around the muon (electron) track, where $\phi$ is the azimuthal angle in radians. 
The sum is corrected for the contribution of neutral particles from additional interactions in the same or nearby bunch crossings (pileup)~\cite{CMS:2018rym,CMS:2015xaf}. 
The relative isolation, defined as the isolation sum divided by the \pt of the lepton, is used in selecting well-isolated electrons and muons. 

Both the SM and BSM analyses use the ``tight'' identification criteria described in Refs.~\cite{CMS:2020uim,CMS:2018rym} to further identify electrons and muons, respectively. 
The electron identification criteria use a boosted decision tree (BDT) trained to identify electrons based on their kinematic values and isolation, whereas muons are required to pass a set of selections, including the requirement that their relative isolation be less than 0.15. 
In addition, a less stringent set of identification and isolation criteria, referred to as the ``veto'' criteria, are used to identify and reject additional leptons.

Jets are clustered using the anti-\kt algorithm~\cite{Cacciari:2011ma,Cacciari:2008gp} taking as input the PF candidates. The clustering algorithm is used with distance parameters of 0.4 and 0.8 to reconstruct so-called AK4 and AK8 jets, respectively. The momenta of jets are corrected to mitigate the contributions of pileup particles, by discarding  particles identified originating from pileup vertices and applying an offset correction to subtract the contributions from neutral particles. Further corrections are derived from simulation and by performing in situ measurements of the momentum balance in data. The jet energy resolution amounts typically to 15--20\% at 30\GeV, 10\% at 100\GeV, and 5\% at 1\TeV~\cite{CMS:2016lmd}. Additional selection criteria are applied to each jet to remove those potentially dominated by anomalous contributions from various subdetector components or reconstruction failures, while jets originating from pileup collisions are rejected via a multivariate discriminator~\cite{CMS:2019kuk}.
The missing transverse momentum vector \ptvecmiss is computed as the negative vector \ptvec sum of all PF candidates reconstructed in the event. Its magnitude is denoted as \ptmiss~\cite{CMS:2019ctu}. The \ptvecmiss is modified to account for corrections to the energy scale of the reconstructed jets in the event. The corrections are propagated via the AK4 jet collection.

The AK4 jets originating from \PQb quark hadronization and decay (\PQb jets) are identified using \DeepJet~\cite{CMS:2017wtu,Bols:2020bkb,CMS-DP-2018-058}, a deep neural network whose inputs are the kinematic values of the jets and their constituents, including secondary vertices associated with the jets. 
A requirement is placed on the output of \DeepJet in order to classify jets as ``\PQb-tagged.'' 
For the chosen working point, the efficiency to select b jets ranges from 71 to 81\%, depending on the \pt and $\eta$ of the jet, and the rate for incorrectly tagging jets originating from the hadronization of gluons or light quarks is approximately 1\%. 
Pairs of \PQb-tagged AK4 jets are used to identify Higgs boson candidates in the SM analysis. 

The AK8 jets are used to reconstruct the Higgs boson candidates in the BSM analysis. 
The mass-decorrelated \Xtobb \ParticleNet tagger~\cite{Qu:2019gqs}, a graph neural network that uses the constituents of an AK8 jet to identify its source, is used to identify \Htobb candidates. 
The soft-drop grooming algorithm \cite{Larkoski:2014wba} with angular exponent $\beta = 0$ and soft radiation fraction $z = 0.1$, also known as the modified mass-drop  algorithm~\cite{Dasgupta:2013ihk}, is applied to the AK8 jets to remove soft and wide-angle radiation when estimating the Higgs boson candidate mass. This estimate of the mass is referred to as the soft-drop mass, or \MSD, whereas the ungroomed jet mass is denoted as $m$. 

\section{Simulated event samples}
We model VBS \WH events at LO using the \MGvATNLO (v2.6.5)~\cite{Alwall:2014hca} Monte Carlo (MC) event generator, interfaced with the \PYTHIA~(v8.240)~\cite{Sjostrand:2014zea} program, which is responsible for the modeling of the boson decays. 
The SM cross section of the generated events is $13 \pm 1$\unit{fb} after the \PW boson decays leptonically and the Higgs boson decays to \bbbar. \PW decays that produce a muon or electron through an intermediate \PGt lepton are included in this sample and considered part of the signal. 
In order to study different \lambdaWZ scenarios, in addition to the SM value of $\kW = \kZ = 1$, we also generate events with ${\kW = -1}$ and ${\kZ = +1}$ to test the hypothesis ${\lambdaWZ = -1}$, by manually modifying the $\kW$ coupling.
These BSM events are generated using a different version (v2.6.1) of \MGvATNLO. With these same settings, a third VBS \WH sample is generated using the \MADGRAPH event-by-event reweighting implementation, with a reweighting scheme for 625 different values of $\kW$ and $\kZ$ in the range $[-2, 2]$, allowing for a two-dimensional scan of the parameter space. 

The main sources of background for both the SM and BSM analyses are \ttbar and \Wjets production, because of the presence of at least one \PW boson and a high jet multiplicity. 
The $\ttbar$ background~\cite{Frixione:2007nw} is generated at next-to-leading-order (NLO) accuracy in perturbative quantum chromodynamics (QCD) with the \POWHEG (v2.0) program~\cite{Nason:2004rx,Frixione:2007vw,Alioli:2010xd}. 
The \Wjets background is generated with \MGvATNLO (v2.6.5) at LO accuracy for up to 4 additional partons merged using the MLM matching scheme~\cite{Alwall:2007fs}.

The largest subleading backgrounds for both the SM and BSM analyses are diboson and single top quark production. 
Diboson events are generated with \MGvATNLO at NLO with up to one additional parton, using the FxFx merging scheme~\cite{Frederix:2012ps}, or with \POWHEG at NLO accuracy. 
The \PQt{}\PW and $t$-channel single top quark processes are generated to NLO accuracy with \POWHEG, whereas $s$-channel single top quark production is modeled with \MGvATNLO (v2.6.5) at NLO accuracy~\cite{Re:2010bp,Frederix:2012dh,Alioli:2009je}.

We also consider backgrounds from VBS production of a \PW boson, quark-initiated \WH, and \Zjets production.
The \Zjets events are generated with the same settings used to generate \Wjets events. 
The VBS \PW background events are also generated with these settings, without any additional jets. 
Finally, quark-initiated \WH production is generated at NLO QCD accuracy using the \POWHEG event generator extended with the MiNLO procedure~\cite{Luisoni:2013cuh}.

The generated matrix-element-level background events are interfaced with \PYTHIA~v8.240 or later versions to model parton showers and hadronization, as well as the underlying event.
In the case of VBS \WH production, we also use \PYTHIA, but with a dipole recoil scheme chosen to correctly emulate the color structure~\cite{Cabouat:2017rzi}.
The generation of all samples is based on the CP5 tune~\cite{tuneCP5Paper} for \PYTHIA and the NNPDF~v3.1 next-to-next-to-leading-order (NNLO) parton distribution functions (PDFs)~\cite{Ball:2014uwa,Ball:2017nwa}.
Simulated pileup interactions are added to the hard-scattering process, with a multiplicity matched to the pileup profiles of the data. 
Finally, the response of the CMS detector is modeled with \GEANTfour \cite{GEANT4:2002zbu}.

\section{Event selection and background estimation}
The selection criteria that target the leptonic \PW boson decay, a single electron or muon plus \ptmiss from a neutrino, are common for the SM and BSM analyses.
Events are required to contain a tight muon with $\pt > 26\,(30)\GeV$ in 2016 and 2018 (2017) data, or a tight electron with $\pt > 30\,(35)\GeV$ in 2016 (2017--18) data. 
Events are rejected if they contain an additional electron or muon with $\pt > 10\GeV$ that satisfies the veto criteria. 
Events are also required to have \ptmiss larger than 30\GeV. 
In addition, in both the SM and BSM analyses, events must contain at least two AK4 jets (the VBS signature jets) that are not \PQb-tagged, do not overlap with the Higgs boson decay products, and satisfy $\pt > 30\GeV$, $\abs{\eta} < 4.7$, $\Mjj > 500\GeV$, and $\absdetajj > 2.5$, where $\Mjj$ is the invariant mass of the dijet system and \detajj is the difference in pseudorapidity between the jets. 

The SM analysis is based on selecting two AK4 jets instead of a single AK8 jet to construct the Higgs boson candidate. 
Specifically, events are required to have two \PQb-tagged AK4 jets with $\pt > 30\GeV$, $\abs{\eta} < 2.5$, and a dijet invariant mass (\Mbb) in the range $50 < \Mbb < 150\GeV$.

A BDT, referred to as the signal BDT, is employed to discriminate between the SM signal and the sum of all backgrounds. The signal events are assigned a target output of 1 and the background events are assigned a target output of 0. The BDT is trained using simulated events to regress the target output with the mean squared error loss function. The training and inference are performed with the XGBoost Python package~\cite{10.1145/2939672.2939785}. The variables that are input to the signal BDT include the \pt, $\eta$, and $\phi$ of the selected lepton and four jets; the \pt and $\phi$ of \ptvecmiss; the \Pb jet discriminator value of the four selected jets; and \Mjj, \detajj, and \Mbb.
Finally, the SM signal region is defined by requiring the signal BDT output to be larger than 0.5, in addition to the \Mbb, \Mjj, and \detajj selections already applied.

The background in the SM BDT output signal region is estimated from two control regions used to determine the overall rate of the \Wjets and \ttbar processes. 
The distributions of these backgrounds, along with both the distributions and overall rates of other smaller backgrounds, are taken from simulation. 
The control region used to constrain the \Wjets process is defined by inverting the BDT output requirement for the signal region. 
In this control region, a second BDT, referred to as the \Wjets BDT, is trained to discriminate between \Wjets and all other processes using the same set of input variables as the signal BDT. 
Then, the \Wjets BDT distribution is used to extract the overall rate of \Wjets production. 
The control region used to constrain the \ttbar process has the signal BDT output requirement removed and the signal region \Mjj requirement modified to $100 < \Mjj < 500\GeV$. 
In this control region, the \Mjj distribution is fit to extract the overall \ttbar production rate.

The signal region for the BSM analysis requires events to have one AK8 jet with $\pt > 300\GeV$, $\abs{\eta} < 2.5$, an \Xtobb \ParticleNet score greater than 0.9, $m > 50\GeV$, and $40 < \MSD < 150\GeV$. 
Events are also required to have no \PQb-tagged AK4 jets. 
Finally, each event must satisfy $\pt(\Pell) + \ptmiss + \pt(\Htobb\text{ jet}) > 900\GeV$, and the requirements on the VBS jets are increased to $\Mjj > 600\GeV$ and $\absdetajj > 4$. 

The background in the BSM signal region is estimated using a modification of the ``ABCD'' method (\eg \cite{CMS:2020cur, CDF:1990kbl}), where two approximately independent variables are used to define four orthogonal regions: region~A, the signal region, and regions B, C, and D, which are dominated by background events. 
In this work, \detajj and \MSD are used, such that inverting one or both selections on each variable gives the background-dominated regions: $\absdetajj \leq 4$ and $\MSD < 150\GeV$ for region B, $\absdetajj > 4$ and $\MSD \geq 150\GeV$ for region C, and $\absdetajj \leq 4$ and $\MSD \geq 150\GeV$ for region D. 
The predicted background yield in the signal region $A_\text{data}^\text{pred}$ is then obtained by the following relation: 
\begin{linenomath*}
\begin{equation*}
    A_\text{data}^\text{pred} = \bigg(B_\text{data}\frac{C_\text{data}}{D_\text{data}}\bigg)\bigg(\frac{A_\mathrm{MC}/B_\mathrm{MC}}{C_\mathrm{MC}/D_\mathrm{MC}}\bigg),
\end{equation*}
\end{linenomath*}
where $B_\text{data}$, $C_\text{data}$, and $D_\text{data}$ are the yields of the three regions in data, and the simulated background yields $A_\mathrm{MC}$, $B_\mathrm{MC}$, $C_\mathrm{MC}$, and $D_\mathrm{MC}$ are used as a correction (right term) to the standard method (left term). 
This correction accounts for a small correlation between \detajj and \MSD for background events that causes an 11\% discrepancy between the ABCD prediction in simulation and $A_\mathrm{MC}$. 
This discrepancy is taken as a baseline for the systematic uncertainty in the prediction.
However, the background composition varies across regions A, B, C, and D, and it is not known to high precision. 
As a result, the baseline systematic uncertainty is extended by varying the subleading backgrounds within their respective uncertainties, representing scenarios in which the contributions from each background are larger or smaller, and recalculating the discrepancy of the method. 

\begin{figure}[htb]
    \centering
    \includegraphics[width=0.49\textwidth]{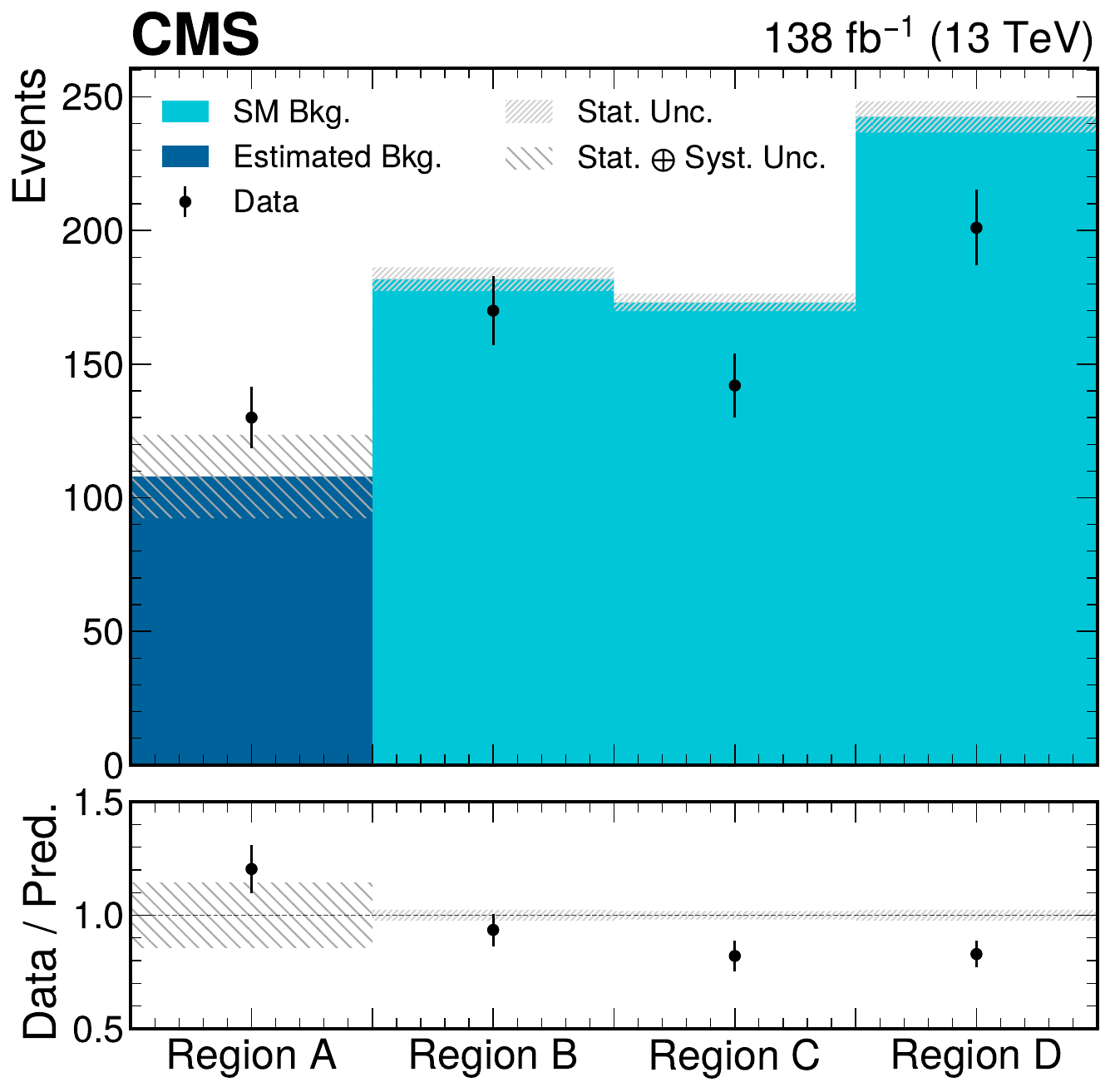}
    \caption{
        Event yields observed in the data (black points) for regions A, B, C, and D of the BSM analysis, where region A is the signal region.     
        The simulated SM background event yields in the control regions B, C, and D are shown in light blue. The background in the signal region estimated from data in regions B, C, and D is plotted in dark blue.
        The lower panel shows the ratio of the data to the estimated background yields. The vertical bars on the points give the statistical uncertainty in the data, while the gray and hatched areas represent the statistical and total uncertainties in the estimated backgrounds, respectively.
    }
    \label{fig:bsm_abcd}
\end{figure}

Based on studies of the discrepancy in the various scenarios, we assign a systematic uncertainty of 13\% in the background prediction.
The yields in the BSM signal region, with $\kW=-1$, $\kZ=1$, are tabulated in Table~\ref{tab:yields} and the yields in regions A, B, C, and D are shown in Fig.~\ref{fig:bsm_abcd}. Another example scenario ($\kW=0$, $\kZ=1$), where the signal sample is generated using the \MADGRAPH event-by-event reweighting, is reported.

\begin{table}[htp]
    \centering
    \topcaption{
	    The background yield estimated from data and the signal yield predicted by MC simulation for the $\kW=-1$, $\kZ=1$ scenario, and for another example scenario ($\kW=0$, $\kZ=1$) in the BSM signal region are shown with their associated statistical and systematic uncertainties. 
        The systematic uncertainty in the signal yield quoted here is the sum in quadrature of the independent relative systematic uncertainties.
        The observed data yield is also tabulated.\\
    }
	{
    \begin{tabular}{lrcc}
        Type       & Yield    & \multicolumn{2}{c}{Uncertainty} \\
                   &     & Stat.  & Syst. \\ 
        \hline
    	VBS WH {($\kappa_{W}=-1$, $\kappa_{Z}=1$)} & 366      & 3          & 68         \\
        VBS WH {($\kappa_{W}=0$, $\kappa_{Z}=1$)}  & 89       & 2          & 17         \\
        Background & 108      & 14         & 14         \\
        Observed data  & 130      &            &            \\
    \end{tabular}
	}
    \label{tab:yields}
\end{table}

\section{Systematic uncertainties}
Results are obtained from binned maximum likelihood fits of the signal and predicted background expectations to the data yields~\cite{Workman:2022ynf}. 
Three distributions---\Mjj in the \ttbar control region, the \Wjets BDT in the \Wjets control region, and the signal BDT in the signal region---are fit simultaneously in the SM analysis, whereas a single-bin counting experiment is performed for the BSM analysis. 
Several sources of systematic uncertainties are estimated for these measurements.
Systematic uncertainties are treated as nuisance parameters with either Gaussian function priors (shape uncertainties) for the SM analysis, or log-normal function priors (normalization uncertainties) for the BSM analysis, and are included in the likelihood function.
Theoretical sources of uncertainty include the renormalization and factorization scales (\muR and \muF, respectively), the PDFs, and the modeling of initial- and final-state radiation. 
Experimental sources of uncertainty include the single-lepton trigger efficiency, lepton identification efficiency, jet energy scale and resolution~\cite{CMS:2016lmd}, \DeepJet \Pb-tagging efficiency, \ParticleNet \Xtobb tagger efficiency, and integrated luminosity measurement~\cite{CMS-LUM-17-003,CMS-PAS-LUM-17-004,CMS-PAS-LUM-18-002}. 
In the SM analysis, the finite statistical power of the simulated samples is considered as a source of uncertainty using the Barlow-Beeston method~\cite{Barlow:1993dm}. 
In the BSM analysis, the statistical uncertainties in the background estimation method and the signal simulation are considered as two additional log-normal constraints. 
The leading systematic uncertainty in the SM signal yield comes from the uncertainty in the \PQb jet tagging efficiency due to the purity of the \PQb jet enriched region where the efficiency is measured in data. 
The rate of non-\PQb jets in this region is subtracted based on simulation, so the uncertainty is determined by varying the size of the subtraction up and down by 20\%~\cite{CMS:2017wtu}. 
The leading systematic uncertainty in the BSM signal yield comes from the variation of \muF. 
In particular, \muF is varied up and down by a factor of 2, and the largest difference in the signal yield is taken as the systematic uncertainty.

\section{Results}
\begin{figure}[htb]
    \centering
    \includegraphics[width=0.49\textwidth]{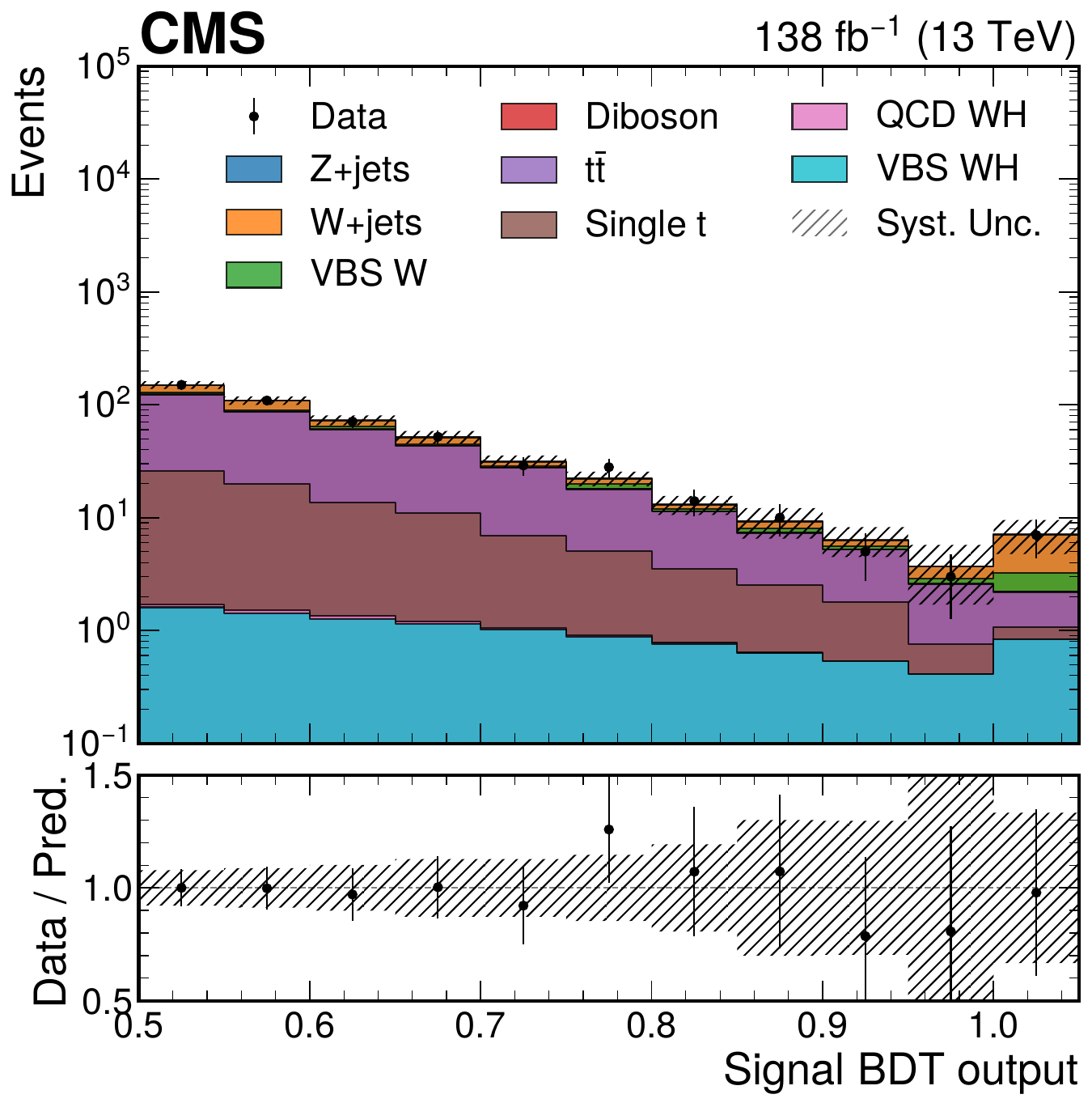}
    \caption{Observed and expected distributions of the BDT output in the signal region of the SM analysis. The data are shown by the points and the expected contributions from the various sources after the fit to the data by the colored histograms.  The lower plot displays the ratio of the data to the sum of the predicted distributions. The vertical bars on the points indicate the statistical uncertainty in the data, and the hatched regions show the systematic uncertainty in the sum of the predicted yields. The rightmost bin includes overflow.}
    \label{fig:resolvedsignalregion}
\end{figure}

The BDT distribution in the signal region for the SM analysis is shown in Fig.~\ref{fig:resolvedsignalregion}. 
The observed (expected) signal strength, the rate of the VBS \WH signal relative to the SM, is $2.2^{+6.1}_{-5.8}\,(1.0^{+4.8}_{-4.1})$. 
The observed (expected) signal significance is 0.34\std (0.22\std), where \std is a standard deviation. 
Finally, the observed (expected) 95\% confidence level CLs~\cite{CLS1,CLS2} upper limit is 14.3 (9.9) times the SM prediction.

\begin{figure}[htb]
    \centering
    \includegraphics[width=0.49\textwidth]{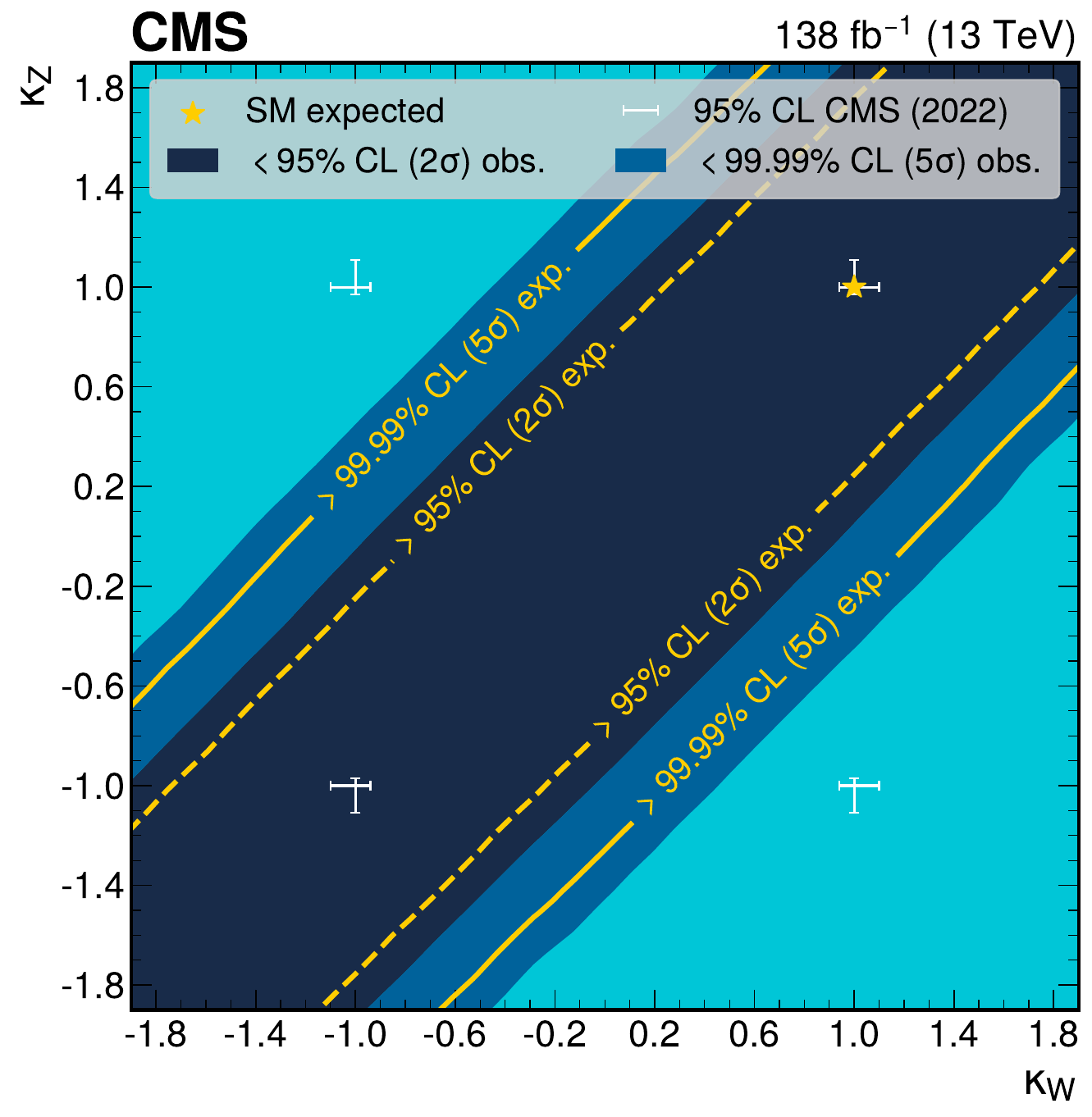}
    \caption{
        The $>$95\% (2\std) and $>$99.99\% (5\std) exclusion regions in the two-dimensional \kW-\kZ plane are shown by the dark and light blue colors, respectively. 
        Previous CMS measurements of the magnitude of \kW and \kZ at the 95\% \CL~\cite{NatureHiggsCMS2022} are represented by the horizontal and vertical bars, and the SM expectation is given by the gold star.
    }
    \label{fig:exclusion2d}
\end{figure}

The limits in the BSM analysis are set following the procedure described in Section~3.2 of Ref.~\cite{Aad2016MLM}. 
We find that the scenario where $\lambdaWZ = \kW/\kZ = -1$, with $\abs{\kW} = 1$ and $\abs{\kZ} = 1$, is excluded at a confidence level (\CL) higher than 99.99\%, with an expected and observed significance beyond 5\std. 
This procedure is repeated for 625 different values of \kW and \kZ.
The exclusion significance regions in the two-dimensional \kW-\kZ plane are shown in Fig.~\ref{fig:exclusion2d}. The observed 95\% \CL (2\std) and 99.99\% (5\std) contours are given by the dark and light blue colors, respectively, while the expected contours are given by the labeled yellow lines. The contours correspond to the conditions $-2\Delta\log{L} < 6.18$ and $<$28.7, respectively, where $L$ is the value of the likelihood function. 
We find that all of the opposite sign scenarios with \kW and \kZ values compatible with the current measurements, $\abs{\kW} = 1.02\pm0.08$ and $\abs{\kZ}= 1.04\pm0.07$~\cite{NatureHiggsCMS2022}, are excluded at a \CL higher than 99.99\%. 

\section{Summary}
In this Letter, we have reported the first study of \WH production through vector boson scattering (VBS) using 138\fbinv of data recorded with the CMS detector at the LHC between 2016 and 2018 at a center-of-mass energy of 13\TeV. 
We focused on both the standard model (SM) scenario and scenarios where $\lambdaWZ=\kW/\kZ$, with \kW and \kZ being the \HWW and \HZZ coupling modifiers, respectively, is modified from the SM value of 1. 
Events were selected by requiring exactly one isolated charged lepton (electron or muon) and the magnitude of the missing transverse momentum \ptmiss being consistent with the \PW boson leptonic decay, two jets consistent with a VBS interaction, and one or two additional jets consistent with the Higgs boson decay to \bbbar.  
Assuming a signal with SM features, the observed (expected) ratio between the measured rate of VBS \WH and the expectation from the SM is $2.2^{+6.1}_{-5.8}\,(1.0^{+4.8}_{-4.1})$, corresponding to an observed (expected) 95\% confidence level upper limit of 14.3 (9.9) times the SM prediction. 
The BSM scenario where $\lambdaWZ = -1$, with $\abs{\kW} = \abs{\kZ} = 1$, and all the opposite-sign scenarios with \kW and \kZ values compatible with the current measurements are excluded with a significance greater than 5 standard deviations, corresponding to a confidence level higher than 99.99\%.

\begin{acknowledgments}
    We congratulate our colleagues in the CERN accelerator departments for the excellent performance of the LHC and thank the technical and administrative staffs at CERN and at other CMS institutes for their contributions to the success of the CMS effort. In addition, we gratefully acknowledge the computing centers and personnel of the Worldwide LHC Computing Grid and other centers for delivering so effectively the computing infrastructure essential to our analyses. Finally, we acknowledge the enduring support for the construction and operation of the LHC, the CMS detector, and the supporting computing infrastructure provided by the following funding agencies: SC (Armenia), BMBWF and FWF (Austria); FNRS and FWO (Belgium); CNPq, CAPES, FAPERJ, FAPERGS, and FAPESP (Brazil); MES and BNSF (Bulgaria); CERN; CAS, MoST, and NSFC (China); MINCIENCIAS (Colombia); MSES and CSF (Croatia); RIF (Cyprus); SENESCYT (Ecuador); ERC PRG, RVTT3 and MoER TK202 (Estonia); Academy of Finland, MEC, and HIP (Finland); CEA and CNRS/IN2P3 (France); SRNSF (Georgia); BMBF, DFG, and HGF (Germany); GSRI (Greece); NKFIH (Hungary); DAE and DST (India); IPM (Iran); SFI (Ireland); INFN (Italy); MSIP and NRF (Republic of Korea); MES (Latvia); LMTLT (Lithuania); MOE and UM (Malaysia); BUAP, CINVESTAV, CONACYT, LNS, SEP, and UASLP-FAI (Mexico); MOS (Montenegro); MBIE (New Zealand); PAEC (Pakistan); MES and NSC (Poland); FCT (Portugal); MESTD (Serbia); MCIN/AEI and PCTI (Spain); MOSTR (Sri Lanka); Swiss Funding Agencies (Switzerland); MST (Taipei); MHESI and NSTDA (Thailand); TUBITAK and TENMAK (Turkey); NASU (Ukraine); STFC (United Kingdom); DOE and NSF (USA).
    
    \hyphenation{Rachada-pisek} Individuals have received support from the Marie-Curie program and the European Research Council and Horizon 2020 Grant, contract Nos.\ 675440, 724704, 752730, 758316, 765710, 824093, 101115353, 101002207, and COST Action CA16108 (European Union); the Leventis Foundation; the Alfred P.\ Sloan Foundation; the Alexander von Humboldt Foundation; the Science Committee, project no. 22rl-037 (Armenia); the Belgian Federal Science Policy Office; the Fonds pour la Formation \`a la Recherche dans l'Industrie et dans l'Agriculture (FRIA-Belgium); the Agentschap voor Innovatie door Wetenschap en Technologie (IWT-Belgium); the F.R.S.-FNRS and FWO (Belgium) under the ``Excellence of Science -- EOS" -- be.h project n.\ 30820817; the Beijing Municipal Science \& Technology Commission, No. Z191100007219010 and Fundamental Research Funds for the Central Universities (China); the Ministry of Education, Youth and Sports (MEYS) of the Czech Republic; the Shota Rustaveli National Science Foundation, grant FR-22-985 (Georgia); the Deutsche Forschungsgemeinschaft (DFG), under Germany's Excellence Strategy -- EXC 2121 ``Quantum Universe" -- 390833306, and under project number 400140256 - GRK2497; the Hellenic Foundation for Research and Innovation (HFRI), Project Number 2288 (Greece); the Hungarian Academy of Sciences, the New National Excellence Program - \'UNKP, the NKFIH research grants K 131991, K 133046, K 138136, K 143460, K 143477, K 146913, K 146914, K 147048, 2020-2.2.1-ED-2021-00181, and TKP2021-NKTA-64 (Hungary); the Council of Science and Industrial Research, India; ICSC -- National Research Center for High Performance Computing, Big Data and Quantum Computing and FAIR -- Future Artificial Intelligence Research, funded by the NextGenerationEU program (Italy); the Latvian Council of Science; the Ministry of Education and Science, project no. 2022/WK/14, and the National Science Center, contracts Opus 2021/41/B/ST2/01369 and 2021/43/B/ST2/01552 (Poland); the Funda\c{c}\~ao para a Ci\^encia e a Tecnologia, grant CEECIND/01334/2018 (Portugal); the National Priorities Research Program by Qatar National Research Fund; MCIN/AEI/10.13039/501100011033, ERDF ``a way of making Europe", and the Programa Estatal de Fomento de la Investigaci{\'o}n Cient{\'i}fica y T{\'e}cnica de Excelencia Mar\'{\i}a de Maeztu, grant MDM-2017-0765 and Programa Severo Ochoa del Principado de Asturias (Spain); the Chulalongkorn Academic into Its 2nd Century Project Advancement Project, and the National Science, Research and Innovation Fund via the Program Management Unit for Human Resources \& Institutional Development, Research and Innovation, grant B37G660013 (Thailand); the Kavli Foundation; the Nvidia Corporation; the SuperMicro Corporation; the Welch Foundation, contract C-1845; and the Weston Havens Foundation (USA).
\end{acknowledgments}

\bibliography{auto_generated}

\providecommand{\href}[2]{#2}\begingroup\raggedright\begin{thebibliography}{10}%
\makeatletter
\providecommand{\hrefCMSnoop }[0]{\@secondoftwo}%
\makeatother
\providecommand{\doi}{\texttt{doi:}\begingroup \urlstyle{tt}\Url}

\bibitem{Englert:1964et}
\hrefCMSnoop {}{F.~Englert and R.~Brout, ``Broken symmetry and the mass of
  gauge vector mesons'',} \textit{ Phys. Rev. Lett.} \textbf{ 13} (1964) 321,
  \href{http://dx.doi.org/10.1103/PhysRevLett.13.321}{\doi{10.1103/PhysRevLett.13.321}}.

\bibitem{Higgs:1964ia}
\hrefCMSnoop {}{P.~W. Higgs, ``Broken symmetries, massless particles and gauge
  fields'',} \textit{ Phys. Lett.} \textbf{ 12} (1964) 132,
  \href{http://dx.doi.org/10.1016/0031-9163(64)91136-9}{\doi{10.1016/0031-9163(64)91136-9}}.

\bibitem{Higgs:1964pj}
\hrefCMSnoop {}{P.~W. Higgs, ``Broken symmetries and the masses of gauge
  bosons'',} \textit{ Phys. Rev. Lett.} \textbf{ 13} (1964) 508,
  \href{http://dx.doi.org/10.1103/PhysRevLett.13.508}{\doi{10.1103/PhysRevLett.13.508}}.

\bibitem{Guralnik:1964eu}
\hrefCMSnoop {}{G.~S. Guralnik, C.~R. Hagen, and T.~W.~B. Kibble, ``Global
  conservation laws and massless particles'',} \textit{ Phys. Rev. Lett.}
  \textbf{ 13} (1964) 585,
  \href{http://dx.doi.org/10.1103/PhysRevLett.13.585}{\doi{10.1103/PhysRevLett.13.585}}.

\bibitem{Higgs:1966ev}
\hrefCMSnoop {}{P.~W. Higgs, ``Spontaneous symmetry breakdown without massless
  bosons'',} \textit{ Phys. Rev.} \textbf{ 145} (1966) 1156,
  \href{http://dx.doi.org/10.1103/PhysRev.145.1156}{\doi{10.1103/PhysRev.145.1156}}.

\bibitem{Kibble:1967sv}
\hrefCMSnoop {}{T.~W.~B. Kibble, ``Symmetry breaking in non-{Abelian} gauge
  theories'',} \textit{ Phys. Rev.} \textbf{ 155} (1967) 1554,
  \href{http://dx.doi.org/10.1103/PhysRev.155.1554}{\doi{10.1103/PhysRev.155.1554}}.

\bibitem{kframe}
\hrefCMSnoop {}{{LHC Higgs Cross Section Working Group}, ``{Higgs} properties:
  Report of the {LHC} {Higgs Cross Section Working Group}'',} in \textit{
  Handbook of LHC Higgs Cross Sections}, CERN Yellow Reports: Monographs.
\newblock 2013.
\newblock \href{http://www.arXiv.org/abs/1307.1347}{\texttt{arXiv:1307.1347}}.

\bibitem{NatureHiggsCMS2022}
\hrefCMSnoop {}{{CMS Collaboration}, ``A portrait of the {Higgs} boson by the
  {CMS} experiment ten years after the discovery'',} \textit{ Nature} \textbf{
  607} (2022) 60,
  \href{http://dx.doi.org/10.1038/s41586-022-04892-x}{\doi{10.1038/s41586-022-04892-x}},
  \href{http://www.arXiv.org/abs/2207.00043}{\texttt{arXiv:2207.00043}}.

\bibitem{NatureHiggsATLAS2022}
\hrefCMSnoop {}{{ATLAS Collaboration}, ``A detailed map of {Higgs} boson
  interactions by the {ATLAS} experiment ten years after the discovery'',}
  \textit{ Nature} \textbf{ 607} (2022) 52,
  \href{http://dx.doi.org/10.1038/s41586-022-04893-w}{\doi{10.1038/s41586-022-04893-w}},
  \href{http://www.arXiv.org/abs/2207.00092}{\texttt{arXiv:2207.00092}}.
  [Erratum: \DOI{10.1038/s41586-022-05581-5}].

\bibitem{Theory2LambdaWZ}
\hrefCMSnoop {}{D.~Stolarski and Y.~Wu, ``Tree-level interference in vector
  boson fusion production of {Vh}'',} \textit{ Phys. Rev. D} \textbf{ 102}
  (2020) 033006,
  \href{http://dx.doi.org/10.1103/PhysRevD.102.033006}{\doi{10.1103/PhysRevD.102.033006}},
  \href{http://www.arXiv.org/abs/2006.09374}{\texttt{arXiv:2006.09374}}.

\bibitem{CMS:2008xjf}
\hrefCMSnoop {}{{CMS Collaboration}, ``The {CMS} experiment at the {CERN}
  {LHC}'',} \textit{ JINST} \textbf{ 3} (2008) S08004,
  \href{http://dx.doi.org/10.1088/1748-0221/3/08/S08004}{\doi{10.1088/1748-0221/3/08/S08004}}.

\bibitem{hepdata}
\hrefCMSnoop {}{}{HEPD}ata record for this analysis, 2024.
\newblock
  \href{http://dx.doi.org/10.17182/hepdata.151278}{\doi{10.17182/hepdata.151278}}.

\bibitem{CMS:2023gfb}
\hrefCMSnoop {}{{CMS Collaboration}, ``Development of the {CMS} detector for
  the {CERN LHC Run 3}'',} \textit{ JINST} \textbf{ 19} (2024) P05064,
  \href{http://dx.doi.org/10.1088/1748-0221/19/05/P05064}{\doi{10.1088/1748-0221/19/05/P05064}},
  \href{http://www.arXiv.org/abs/2309.05466}{\texttt{arXiv:2309.05466}}.

\bibitem{CMS:2020cmk}
\hrefCMSnoop {}{{CMS Collaboration}, ``Performance of the {CMS} {Level-1}
  trigger in proton-proton collisions at $\sqrt{s} = 13$ {TeV}'',} \textit{
  JINST} \textbf{ 15} (2020) P10017,
  \href{http://dx.doi.org/10.1088/1748-0221/15/10/P10017}{\doi{10.1088/1748-0221/15/10/P10017}},
  \href{http://www.arXiv.org/abs/2006.10165}{\texttt{arXiv:2006.10165}}.

\bibitem{CMS:2016ngn}
\hrefCMSnoop {}{{CMS Collaboration}, ``The {CMS} trigger system'',} \textit{
  JINST} \textbf{ 12} (2017) P01020,
  \href{http://dx.doi.org/10.1088/1748-0221/12/01/P01020}{\doi{10.1088/1748-0221/12/01/P01020}},
\href{http://www.arXiv.org/abs/1609.02366}{\texttt{arXiv:1609.02366}}.
%%CITATION = ARXIV:1609.02366;%%.

\bibitem{CMS:2020uim}
\hrefCMSnoop {}{{CMS Collaboration}, ``Electron and photon reconstruction and
  identification with the {CMS} experiment at the {CERN} {LHC}'',} \textit{
  JINST} \textbf{ 16} (2021) P05014,
  \href{http://dx.doi.org/10.1088/1748-0221/16/05/P05014}{\doi{10.1088/1748-0221/16/05/P05014}},
  \href{http://www.arXiv.org/abs/2012.06888}{\texttt{arXiv:2012.06888}}.

\bibitem{CMS:2018rym}
\hrefCMSnoop {}{{CMS Collaboration}, ``Performance of the {CMS} muon detector
  and muon reconstruction with proton-proton collisions at $\sqrt{s}=13$
  {TeV}'',} \textit{ JINST} \textbf{ 13} (2018) P06015,
  \href{http://dx.doi.org/10.1088/1748-0221/13/06/P06015}{\doi{10.1088/1748-0221/13/06/P06015}},
  \href{http://www.arXiv.org/abs/1804.04528}{\texttt{arXiv:1804.04528}}.

\bibitem{CMS:2014pgm}
\hrefCMSnoop {}{{CMS Collaboration}, ``Description and performance of track and
  primary-vertex reconstruction with the {CMS} tracker'',} \textit{ JINST}
  \textbf{ 9} (2014) P10009,
  \href{http://dx.doi.org/10.1088/1748-0221/9/10/P10009}{\doi{10.1088/1748-0221/9/10/P10009}},
  \href{http://www.arXiv.org/abs/1405.6569}{\texttt{arXiv:1405.6569}}.

\bibitem{CMS:2017yfk}
\hrefCMSnoop {}{{CMS Collaboration}, ``Particle-flow reconstruction and global
  event description with the {CMS} detector'',} \textit{ JINST} \textbf{ 12}
  (2017) P10003,
  \href{http://dx.doi.org/10.1088/1748-0221/12/10/P10003}{\doi{10.1088/1748-0221/12/10/P10003}},
\href{http://www.arXiv.org/abs/1706.04965}{\texttt{arXiv:1706.04965}}.
%%CITATION = ARXIV:1706.04965;%%.

\bibitem{CMS:2018jrd}
\hrefCMSnoop {}{{CMS Collaboration}, ``Performance of reconstruction and
  identification of $\tau$ leptons decaying to hadrons and $\nu_\tau$ in pp
  collisions at $\sqrt{s} = 13$ {TeV}'',} \textit{ JINST} \textbf{ 13} (2018)
  P10005,
  \href{http://dx.doi.org/10.1088/1748-0221/13/10/P10005}{\doi{10.1088/1748-0221/13/10/P10005}},
  \href{http://www.arXiv.org/abs/1809.02816}{\texttt{arXiv:1809.02816}}.

\bibitem{CMS:2016lmd}
\hrefCMSnoop {}{{CMS Collaboration}, ``Jet energy scale and resolution in the
  {CMS} experiment in pp collisions at 8 {TeV}'',} \textit{ JINST} \textbf{ 12}
  (2017) P02014,
  \href{http://dx.doi.org/10.1088/1748-0221/12/02/P02014}{\doi{10.1088/1748-0221/12/02/P02014}},
\href{http://www.arXiv.org/abs/1607.03663}{\texttt{arXiv:1607.03663}}.
%%CITATION = ARXIV:1607.03663;%%.

\bibitem{CMS:2019ctu}
\hrefCMSnoop {}{{CMS Collaboration}, ``Performance of missing transverse
  momentum reconstruction in proton-proton collisions at $\sqrt{s} =13$ {TeV}
  using the {CMS} detector'',} \textit{ JINST} \textbf{ 14} (2019) P07004,
  \href{http://dx.doi.org/10.1088/1748-0221/14/07/P07004}{\doi{10.1088/1748-0221/14/07/P07004}},
  \href{http://www.arXiv.org/abs/1903.06078}{\texttt{arXiv:1903.06078}}.

\bibitem{CMS-TDR-15-02}
\href {http://cds.cern.ch/record/2020886}{{CMS Collaboration}, ``Technical
  proposal for the {Phase-II} upgrade of the {Compact Muon Solenoid}'',} CMS
  Technical Proposal CERN-LHCC-2015-010, CMS-TDR-15-02, 2015.

\bibitem{CMS:2015xaf}
\hrefCMSnoop {}{{CMS Collaboration}, ``Performance of electron reconstruction
  and selection with the {CMS} detector in proton-proton collisions at
  $\sqrt{s}=8$ {TeV}'',} \textit{ JINST} \textbf{ 10} (2015) P06005,
  \href{http://dx.doi.org/10.1088/1748-0221/10/06/P06005}{\doi{10.1088/1748-0221/10/06/P06005}},
  \href{http://www.arXiv.org/abs/1502.02701}{\texttt{arXiv:1502.02701}}.

\bibitem{Cacciari:2011ma}
\hrefCMSnoop {}{M.~Cacciari, G.~P. Salam, and G.~Soyez, ``{FastJet} user
  manual'',} \textit{ Eur. Phys. J. C} \textbf{ 72} (2012) 1896,
  \href{http://dx.doi.org/10.1140/epjc/s10052-012-1896-2}{\doi{10.1140/epjc/s10052-012-1896-2}},
  \href{http://www.arXiv.org/abs/1111.6097}{\texttt{arXiv:1111.6097}}.

\bibitem{Cacciari:2008gp}
\hrefCMSnoop {}{M.~Cacciari, G.~P. Salam, and G.~Soyez, ``The anti-\kt jet
  clustering algorithm'',} \textit{ JHEP} \textbf{ 04} (2008) 063,
  \href{http://dx.doi.org/10.1088/1126-6708/2008/04/063}{\doi{10.1088/1126-6708/2008/04/063}},
  \href{http://www.arXiv.org/abs/0802.1189}{\texttt{arXiv:0802.1189}}.

\bibitem{CMS:2019kuk}
\hrefCMSnoop {}{{CMS Collaboration}, ``Pileup mitigation at {CMS} in 13 {TeV}
  data'',} \textit{ JINST} \textbf{ 15} (2020) P09018,
  \href{http://dx.doi.org/10.1088/1748-0221/15/09/P09018}{\doi{10.1088/1748-0221/15/09/P09018}},
  \href{http://www.arXiv.org/abs/2003.00503}{\texttt{arXiv:2003.00503}}.

\bibitem{CMS:2017wtu}
\hrefCMSnoop {}{{CMS Collaboration}, ``Identification of heavy-flavour jets
  with the {CMS} detector in pp collisions at 13 {TeV}'',} \textit{ JINST}
  \textbf{ 13} (2018) P05011,
  \href{http://dx.doi.org/10.1088/1748-0221/13/05/P05011}{\doi{10.1088/1748-0221/13/05/P05011}},
  \href{http://www.arXiv.org/abs/1712.07158}{\texttt{arXiv:1712.07158}}.

\bibitem{Bols:2020bkb}
E.~Bols\hrefCMSnoop {}{ { et~al.}, ``Jet flavour classification using
  {DeepJet}'',} \textit{ JINST} \textbf{ 15} (2020) P12012,
  \href{http://dx.doi.org/10.1088/1748-0221/15/12/P12012}{\doi{10.1088/1748-0221/15/12/P12012}},
  \href{http://www.arXiv.org/abs/2008.10519}{\texttt{arXiv:2008.10519}}.

\bibitem{CMS-DP-2018-058}
\href {http://cds.cern.ch/record/2646773}{{CMS Collaboration}, ``Performance of
  the {DeepJet} b tagging algorithm using 41.9/fb of data from proton-proton
  collisions at 13 {TeV} with phase 1 {CMS} detector'',} {CMS} Detector
  Performance Summary CMS-DP-2018-058, 2018.

\bibitem{Qu:2019gqs}
\hrefCMSnoop {}{H.~Qu and L.~Gouskos, ``{ParticleNet}: Jet tagging via particle
  clouds'',} \textit{ Phys. Rev. D} \textbf{ 101} (2020) 056019,
  \href{http://dx.doi.org/10.1103/PhysRevD.101.056019}{\doi{10.1103/PhysRevD.101.056019}},
  \href{http://www.arXiv.org/abs/1902.08570}{\texttt{arXiv:1902.08570}}.

\bibitem{Larkoski:2014wba}
\hrefCMSnoop {}{A.~J. Larkoski, S.~Marzani, G.~Soyez, and J.~Thaler, ``Soft
  drop'',} \textit{ JHEP} \textbf{ 05} (2014) 146,
  \href{http://dx.doi.org/10.1007/JHEP05(2014)146}{\doi{10.1007/JHEP05(2014)146}},
  \href{http://www.arXiv.org/abs/1402.2657}{\texttt{arXiv:1402.2657}}.

\bibitem{Dasgupta:2013ihk}
\hrefCMSnoop {}{M.~Dasgupta, A.~Fregoso, S.~Marzani, and G.~P. Salam, ``Towards
  an understanding of jet substructure'',} \textit{ JHEP} \textbf{ 09} (2013)
  029,
  \href{http://dx.doi.org/10.1007/JHEP09(2013)029}{\doi{10.1007/JHEP09(2013)029}},
  \href{http://www.arXiv.org/abs/1307.0007}{\texttt{arXiv:1307.0007}}.

\bibitem{Alwall:2014hca}
J.~Alwall\hrefCMSnoop {}{ { et~al.}, ``The automated computation of tree-level
  and next-to-leading order differential cross sections, and their matching to
  parton shower simulations'',} \textit{ JHEP} \textbf{ 07} (2014) 079,
  \href{http://dx.doi.org/10.1007/JHEP07(2014)079}{\doi{10.1007/JHEP07(2014)079}},
  \href{http://www.arXiv.org/abs/1405.0301}{\texttt{arXiv:1405.0301}}.

\bibitem{Sjostrand:2014zea}
T.~Sj{\"o}strand\hrefCMSnoop {}{ { et~al.}, ``An introduction to {PYTHIA}
  8.2'',} \textit{ Comput. Phys. Commun.} \textbf{ 191} (2015) 159,
  \href{http://dx.doi.org/10.1016/j.cpc.2015.01.024}{\doi{10.1016/j.cpc.2015.01.024}},
\href{http://www.arXiv.org/abs/1410.3012}{\texttt{arXiv:1410.3012}}.
%%CITATION = ARXIV:1410.3012;%%.

\bibitem{Frixione:2007nw}
\hrefCMSnoop {}{S.~Frixione, P.~Nason, and G.~Ridolfi, ``A positive-weight
  next-to-leading-order {Monte Carlo} for heavy flavour hadroproduction'',}
  \textit{ JHEP} \textbf{ 09} (2007) 126,
  \href{http://dx.doi.org/10.1088/1126-6708/2007/09/126}{\doi{10.1088/1126-6708/2007/09/126}},
  \href{http://www.arXiv.org/abs/0707.3088}{\texttt{arXiv:0707.3088}}.

\bibitem{Nason:2004rx}
\hrefCMSnoop {}{P.~Nason, ``A new method for combining {NLO} {QCD} with shower
  {Monte Carlo} algorithms'',} \textit{ JHEP} \textbf{ 11} (2004) 040,
  \href{http://dx.doi.org/10.1088/1126-6708/2004/11/040}{\doi{10.1088/1126-6708/2004/11/040}},
\href{http://www.arXiv.org/abs/hep-ph/0409146}{\texttt{arXiv:hep-ph/0409146}}.
%%CITATION = HEP-PH/0409146;%%.

\bibitem{Frixione:2007vw}
\hrefCMSnoop {}{S.~Frixione, P.~Nason, and C.~Oleari, ``Matching {NLO} {QCD}
  computations with parton shower simulations: the {POWHEG} method'',} \textit{
  JHEP} \textbf{ 11} (2007) 070,
  \href{http://dx.doi.org/10.1088/1126-6708/2007/11/070}{\doi{10.1088/1126-6708/2007/11/070}},
\href{http://www.arXiv.org/abs/0709.2092}{\texttt{arXiv:0709.2092}}.
%%CITATION = ARXIV:0709.2092;%%.

\bibitem{Alioli:2010xd}
\hrefCMSnoop {}{S.~Alioli, P.~Nason, C.~Oleari, and E.~Re, ``A general
  framework for implementing {NLO} calculations in shower {Monte Carlo}
  programs: the {POWHEG BOX}'',} \textit{ JHEP} \textbf{ 06} (2010) 043,
  \href{http://dx.doi.org/10.1007/JHEP06(2010)043}{\doi{10.1007/JHEP06(2010)043}},
\href{http://www.arXiv.org/abs/1002.2581}{\texttt{arXiv:1002.2581}}.
%%CITATION = ARXIV:1002.2581;%%.

\bibitem{Alwall:2007fs}
J.~Alwall\hrefCMSnoop {}{ { et~al.}, ``Comparative study of various algorithms
  for the merging of parton showers and matrix elements in hadronic
  collisions'',} \textit{ Eur. Phys. J. C} \textbf{ 53} (2008) 473,
  \href{http://dx.doi.org/10.1140/epjc/s10052-007-0490-5}{\doi{10.1140/epjc/s10052-007-0490-5}},
\href{http://www.arXiv.org/abs/0706.2569}{\texttt{arXiv:0706.2569}}.
%%CITATION = ARXIV:0706.2569;%%.

\bibitem{Frederix:2012ps}
\hrefCMSnoop {}{R.~Frederix and S.~Frixione, ``Merging meets matching in
  {MC@NLO}'',} \textit{ JHEP} \textbf{ 12} (2012) 061,
  \href{http://dx.doi.org/10.1007/JHEP12(2012)061}{\doi{10.1007/JHEP12(2012)061}},
\href{http://www.arXiv.org/abs/1209.6215}{\texttt{arXiv:1209.6215}}.
%%CITATION = ARXIV:1209.6215;%%.

\bibitem{Re:2010bp}
\hrefCMSnoop {}{E.~Re, ``Single-top {Wt-channel} production matched with parton
  showers using the {POWHEG} method'',} \textit{ Eur. Phys. J. C} \textbf{ 71}
  (2011) 1547,
  \href{http://dx.doi.org/10.1140/epjc/s10052-011-1547-z}{\doi{10.1140/epjc/s10052-011-1547-z}},
  \href{http://www.arXiv.org/abs/1009.2450}{\texttt{arXiv:1009.2450}}.

\bibitem{Frederix:2012dh}
\hrefCMSnoop {}{R.~Frederix, E.~Re, and P.~Torrielli, ``Single-top $t$-channel
  hadroproduction in the four-flavour scheme with {POWHEG} and {aMC@NLO}'',}
  \textit{ JHEP} \textbf{ 09} (2012) 130,
  \href{http://dx.doi.org/10.1007/JHEP09(2012)130}{\doi{10.1007/JHEP09(2012)130}},
  \href{http://www.arXiv.org/abs/1207.5391}{\texttt{arXiv:1207.5391}}.

\bibitem{Alioli:2009je}
\hrefCMSnoop {}{S.~Alioli, P.~Nason, C.~Oleari, and E.~Re, ``{NLO} single-top
  production matched with shower in {POWHEG}: $s$- and $t$-channel
  contributions'',} \textit{ JHEP} \textbf{ 09} (2009) 111,
  \href{http://dx.doi.org/10.1088/1126-6708/2009/09/111}{\doi{10.1088/1126-6708/2009/09/111}},
  \href{http://www.arXiv.org/abs/0907.4076}{\texttt{arXiv:0907.4076}}.
  [Erratum: \DOI{10.1007/JHEP02(2010)011}].

\bibitem{Luisoni:2013cuh}
\hrefCMSnoop {}{G.~Luisoni, P.~Nason, C.~Oleari, and F.~Tramontano,
  ``{HW}$^{\pm}$/{HZ} + 0 and 1 jet at {NLO} with the {POWHEG} {BOX} interfaced
  to {GoSam} and their merging within {MiNLO}'',} \textit{ JHEP} \textbf{ 10}
  (2013) 083,
  \href{http://dx.doi.org/10.1007/JHEP10(2013)083}{\doi{10.1007/JHEP10(2013)083}},
  \href{http://www.arXiv.org/abs/1306.2542}{\texttt{arXiv:1306.2542}}.

\bibitem{Cabouat:2017rzi}
\hrefCMSnoop {}{B.~Cabouat and T.~Sj{\"o}strand, ``Some dipole shower
  studies'',} \textit{ Eur. Phys. J. C} \textbf{ 78} (2018) 226,
  \href{http://dx.doi.org/10.1140/epjc/s10052-018-5645-z}{\doi{10.1140/epjc/s10052-018-5645-z}},
  \href{http://www.arXiv.org/abs/1710.00391}{\texttt{arXiv:1710.00391}}.

\bibitem{tuneCP5Paper}
\hrefCMSnoop {}{{CMS Collaboration}, ``Extraction and validation of a new set
  of {CMS} {PYTHIA8} tunes from underlying-event measurements'',} \textit{ Eur.
  Phys. J. C} \textbf{ 80} (2020) 4,
  \href{http://dx.doi.org/10.1140/epjc/s10052-019-7499-4}{\doi{10.1140/epjc/s10052-019-7499-4}},
  \href{http://www.arXiv.org/abs/1903.12179}{\texttt{arXiv:1903.12179}}.

\bibitem{Ball:2014uwa}
\hrefCMSnoop {}{{NNPDF} Collaboration, ``Parton distributions for the {LHC}
  {Run II}'',} \textit{ JHEP} \textbf{ 04} (2015) 040,
  \href{http://dx.doi.org/10.1007/JHEP04(2015)040}{\doi{10.1007/JHEP04(2015)040}},
\href{http://www.arXiv.org/abs/1410.8849}{\texttt{arXiv:1410.8849}}.
%%CITATION = ARXIV:1410.8849;%%.

\bibitem{Ball:2017nwa}
\hrefCMSnoop {}{{NNPDF} Collaboration, ``Parton distributions from
  high-precision collider data'',} \textit{ Eur. Phys. J. C} \textbf{ 77}
  (2017) 663,
  \href{http://dx.doi.org/10.1140/epjc/s10052-017-5199-5}{\doi{10.1140/epjc/s10052-017-5199-5}},
\href{http://www.arXiv.org/abs/1706.00428}{\texttt{arXiv:1706.00428}}.
%%CITATION = ARXIV:1706.00428;%%.

\bibitem{GEANT4:2002zbu}
\hrefCMSnoop {}{{GEANT4} Collaboration, ``{\GEANTfour}---a simulation
  toolkit'',} \textit{ Nucl. Instrum. Meth. A} \textbf{ 506} (2003) 250,
\href{http://dx.doi.org/10.1016/S0168-9002(03)01368-8}{\doi{10.1016/S0168-9002(03)01368-8}}.
%%CITATION = NUIMA,A506,250;%%.

\bibitem{10.1145/2939672.2939785}
\hrefCMSnoop {}{T.~Chen and C.~Guestrin, ``{XGBoost}: A scalable tree boosting
  system'',} in \textit{ Proc. 22nd ACM SIGKDD Int. Conf. on Knowledge
  Discovery and Data Mining}, KDD '16, p.~785.
\newblock Association for Computing Machinery, New York, NY, USA, 2016.
\newblock
  \href{http://dx.doi.org/10.1145/2939672.2939785}{\doi{10.1145/2939672.2939785}}.

\bibitem{CMS:2020cur}
\hrefCMSnoop {}{{CMS Collaboration}, ``Search for supersymmetry in pp
  collisions at $\sqrt{s} = 13$ {TeV} with 137 fb$^{-1}$ in final states with a
  single lepton using the sum of masses of large-radius jets'',} \textit{ Phys.
  Rev. D} \textbf{ 101} (2020) 052010,
  \href{http://dx.doi.org/10.1103/PhysRevD.101.052010}{\doi{10.1103/PhysRevD.101.052010}},
  \href{http://www.arXiv.org/abs/1911.07558}{\texttt{arXiv:1911.07558}}.

\bibitem{CDF:1990kbl}
\hrefCMSnoop {}{{CDF} Collaboration, ``A measurement of $\sigma {B} ({W} \to e
  \nu)$ and $\sigma {B} ({Z}^0 \to e^+ e-)$ in $\bar{p}p$ collisions at
  $\sqrt{s} = 1800$ {GeV}'',} \textit{ Phys. Rev. D} \textbf{ 44} (1991) 29,
  \href{http://dx.doi.org/10.1103/PhysRevD.44.29}{\doi{10.1103/PhysRevD.44.29}}.

\bibitem{Workman:2022ynf}
\hrefCMSnoop {}{{Particle Data Group}, R.~L. Workman { et~al.}, ``Review of
  particle physics'',} \textit{ Prog. Theor. Exp. Phys.} \textbf{ 2022} (2022)
  083C01,
  \href{http://dx.doi.org/10.1093/ptep/ptac097}{\doi{10.1093/ptep/ptac097}}.

\bibitem{CMS-LUM-17-003}
\hrefCMSnoop {}{{CMS Collaboration}, ``Precision luminosity measurement in
  proton-proton collisions at {$\sqrt{s} = 13\TeV$} in 2015 and 2016 at
  {CMS}'',} \textit{ Eur. Phys. J. C} \textbf{ 81} (2021) 800,
  \href{http://dx.doi.org/10.1140/epjc/s10052-021-09538-2}{\doi{10.1140/epjc/s10052-021-09538-2}},
  \href{http://www.arXiv.org/abs/2104.01927}{\texttt{arXiv:2104.01927}}.

\bibitem{CMS-PAS-LUM-17-004}
\href {https://cds.cern.ch/record/2621960/}{{CMS Collaboration}, ``{CMS}
  luminosity measurement for the 2017 data-taking period at {$\sqrt{s} =
  13\TeV$}'',} {CMS Physics Analysis Summary}, 2018.

\bibitem{CMS-PAS-LUM-18-002}
\href {https://cds.cern.ch/record/2676164/}{{CMS Collaboration}, ``{CMS}
  luminosity measurement for the 2018 data-taking period at {$\sqrt{s} =
  13\TeV$}'',} {CMS Physics Analysis Summary}, 2019.

\bibitem{Barlow:1993dm}
\hrefCMSnoop {}{R.~J. Barlow and C.~Beeston, ``Fitting using finite {Monte
  Carlo} samples'',} \textit{ Comput. Phys. Commun.} \textbf{ 77} (1993) 219,
  \href{http://dx.doi.org/10.1016/0010-4655(93)90005-W}{\doi{10.1016/0010-4655(93)90005-W}}.

\bibitem{CLS1}
\hrefCMSnoop {}{A.~L. Read, ``Presentation of search results: The
  {CL$_{\text{s}}$} technique'',} \textit{ J. Phys. G} \textbf{ 28} (2002)
  2693,
\href{http://dx.doi.org/10.1088/0954-3899/28/10/313}{\doi{10.1088/0954-3899/28/10/313}}.
%%CITATION = JPAGA,G28,2693;%%.

\bibitem{CLS2}
\hrefCMSnoop {}{T.~Junk, ``Confidence level computation for combining searches
  with small statistics'',} \textit{ Nucl. Instrum. Meth. A} \textbf{ 434}
  (1999) 435,
  \href{http://dx.doi.org/10.1016/S0168-9002(99)00498-2}{\doi{10.1016/S0168-9002(99)00498-2}},
\href{http://www.arXiv.org/abs/hep-ex/9902006}{\texttt{arXiv:hep-ex/9902006}}.
%%CITATION = HEP-EX/9902006;%%.

\bibitem{Aad2016MLM}
\hrefCMSnoop {}{{ATLAS and CMS Collaborations}, ``Measurements of the {Higgs}
  boson production and decay rates and constraints on its couplings from a
  combined {ATLAS} and {CMS} analysis of the {LHC} pp collision data at $
  \sqrt{s}=7 $ and 8 {TeV}'',} \textit{ JHEP} \textbf{ 08} (2016) 045,
  \href{http://dx.doi.org/10.1007/JHEP08(2016)045}{\doi{10.1007/JHEP08(2016)045}},
  \href{http://www.arXiv.org/abs/1606.02266}{\texttt{arXiv:1606.02266}}.

\end{thebibliography}\endgroup
\cleardoublepage \appendix\section{The CMS Collaboration \label{app:collab}}\begin{sloppypar}\hyphenpenalty=5000\widowpenalty=500\clubpenalty=5000
\cmsinstitute{Yerevan Physics Institute, Yerevan, Armenia}
{\tolerance=6000
A.~Hayrapetyan, A.~Tumasyan\cmsAuthorMark{1}\cmsorcid{0009-0000-0684-6742}
\par}
\cmsinstitute{Institut f\"{u}r Hochenergiephysik, Vienna, Austria}
{\tolerance=6000
W.~Adam\cmsorcid{0000-0001-9099-4341}, J.W.~Andrejkovic, T.~Bergauer\cmsorcid{0000-0002-5786-0293}, S.~Chatterjee\cmsorcid{0000-0003-2660-0349}, K.~Damanakis\cmsorcid{0000-0001-5389-2872}, M.~Dragicevic\cmsorcid{0000-0003-1967-6783}, P.S.~Hussain\cmsorcid{0000-0002-4825-5278}, M.~Jeitler\cmsAuthorMark{2}\cmsorcid{0000-0002-5141-9560}, N.~Krammer\cmsorcid{0000-0002-0548-0985}, A.~Li\cmsorcid{0000-0002-4547-116X}, D.~Liko\cmsorcid{0000-0002-3380-473X}, I.~Mikulec\cmsorcid{0000-0003-0385-2746}, J.~Schieck\cmsAuthorMark{2}\cmsorcid{0000-0002-1058-8093}, R.~Sch\"{o}fbeck\cmsorcid{0000-0002-2332-8784}, D.~Schwarz\cmsorcid{0000-0002-3821-7331}, M.~Sonawane\cmsorcid{0000-0003-0510-7010}, S.~Templ\cmsorcid{0000-0003-3137-5692}, W.~Waltenberger\cmsorcid{0000-0002-6215-7228}, C.-E.~Wulz\cmsAuthorMark{2}\cmsorcid{0000-0001-9226-5812}
% ...
\par}
\vskip\cmsinstskip
\dag:~Deceased\\
$^{1}$Also at Yerevan State University, Yerevan, Armenia\\
$^{2}$Also at TU Wien, Vienna, Austria\\
% ...
\end{sloppypar}
%%% END EDITABLE REGION %%%
% skeleton_end
\end{document}